\documentclass{article}

\usepackage{psfig}
\usepackage{emulateapj}
\usepackage{apjfonts}

\makeatletter
\let\@internalcite\cite
\def\cite{\def\astroncite##1##2{##1\ ##2}\@internalcite}
\def\citey{\def\astroncite##1##2{##1\ (##2)}\@internalcite}
\def\@citex[#1]#2{\if@filesw\immediate\write\@auxout{\string\citation{#2}}\fi
  \def\@citea{}\@cite{\@for\@citeb:=#2\do
    {\@citea\def\@citea{; }\@ifundefined
       {b@\@citeb}{{\bf ??}\@warning
       {Citation `\@citeb' on page \thepage \space undefined}}%
{\csname b@\@citeb\endcsname}}}{#1}}
\def\@cite#1#2{#1\if@tempswa #2\fi}
\def\@biblabel#1{}
\def\astroncite#1#2{#1\ #2}
\makeatother

\def\cm{{\rm cm}}

\def\aproxgt{\mathrel{%
      \rlap{\raise 0.511ex \hbox{$>$}}{\lower 0.511ex \hbox{$\sim$}}}}
\def\aproxlt{\mathrel{%
      \rlap{\raise 0.511ex \hbox{$<$}}{\lower 0.511ex \hbox{$\sim$}}}}

\newcommand{\Ka}{K$\alpha$}
\newcommand{\Kb}{K$\beta$}
\newcommand{\Ne}{\ensuremath{N_{\mbox{\scriptsize\sc e}}}}
\newcommand{\me}{m e}
\def\keV{\rm keV}
\def\cyg{Cyg X--1 }
\def\be{\begin{equation}}
\def\ee{\end{equation}}
\def\bea{\begin{eqnarray}}
\def\eea{\end{eqnarray}}
\def\gamfn#1{ { \Gamma \left ( #1 \right ) } }

\begin{document}

\slugcomment{To Appear in The Astrophysical Journal, 1999}

\lefthead{Nowak et al.}
\righthead{RXTE Observation of Cygnus X--1: III.}

\title{RXTE Observation of Cygnus X--1: III. Implications for Compton Corona
and ADAF Models}

\author{Michael A. Nowak\altaffilmark{1}, J\"orn Wilms\altaffilmark{2},
Brian A. Vaughan\altaffilmark{3}, James B. Dove\altaffilmark{1,4}, Mitchell
C. Begelman\altaffilmark{1,5}} 

\altaffiltext{1}{JILA, University of Colorado, Campus Box 440, Boulder, CO
  ~80309-0440; \{mnowak, dove\}@rocinante.colorado.edu,
  mitch@jila.colorado.edu} 
\altaffiltext{2}{Institut f\"ur Astronomie und
  Astrophysik, Abt.~Astronomie, Waldh\"auser Str. 64, D-72076 T\"ubingen,
  Germany; wilms@astro.uni-tuebingen.de} 
\altaffiltext{3}{Space Radiation
  Laboratory, California Institute of Technology, MC 220-47, Pasadena, CA
  ~91125, USA; ~brian@srl.caltech.edu} 
\altaffiltext{4}{also, CASA,
  University of Colorado, Campus Box 389, Boulder, CO ~80309-0389}
\altaffiltext{5}{also, Department of Astronomy, University of Colorado,
  Boulder, CO ~80309, USA}

\received{May 27, 1998}
\accepted{October 24, 1998}

\begin{abstract}
  
  We have recently shown that a `sphere+disk' geometry Compton corona model
  provides a good description of Rossi X-ray Timing Explorer (RXTE)
  observations of the hard/low state of Cygnus X--1.  Separately, we have
  analyzed the temporal data provided by RXTE.  In this paper we consider
  the implications of this timing analysis for our best-fit
  `sphere+disk' Comptonization models.  We focus our attention on the observed
  Fourier frequency-dependent time delays between hard and soft photons.
  We consider whether the observed time delays are: created in the disk but
  are merely reprocessed by the corona; created by differences between the
  hard and soft photon diffusion times in coronae with extremely large
  radii; or are due to `propagation' of disturbances through the corona.
  We find that the time delays are most likely created directly within the
  corona; however, it is currently uncertain which specific model is the
  most likely explanation.  Models that posit a large coronal radius [or
  equivalently, a large Advection Dominated Accretion Flow (ADAF) region]
  do not fully address all the details of the observed spectrum.  The
  Compton corona models that do address the full spectrum do not contain
  dynamical information.  We show, however, that simple phenomenological
  propagation models for the observed time delays for these latter models
  imply \emph{extremely} slow characteristic propagation speeds within the
  coronal region.

\end{abstract}


\setcounter{footnote}{0}

\section{Introduction}\label{sec:intro}

In a previous paper (\cite{dove:98a}, hereafter paper I) and in a companion
paper to this work (\cite{nowak:98a}, hereafter paper II) we have presented
analysis of a 20\,ksec Rossi X-ray Timing Explorer (RXTE) observation of the
black hole candidate Cygnus~X--1.  Using self-consistent numerical models
of a hot spherical corona surrounded by a cold, geometrically thin disk, we
were able to describe successfully the spectrum of \cyg over a broad range
in energy, $3$--$200$\,keV (paper I; see also \cite{dove:97b}).  We
derived an optical depth for the spherical corona of $\tau=1.6\pm 0.1$ and
an average temperature of $kT = 87 \pm 5$\,keV (reduced $\chi^2$ for the
fit was 1.56; paper I).

Timing analysis (paper II) showed that our observation of the hard state of
\cyg was similar to previous hard state observations of this object
(\cite{miyamoto:89a,belloni:90a,belloni:90b,miyamoto:92a}); however, we
were able to extend our analysis to a decade lower in Fourier frequency and
half a decade higher in Fourier frequency as compared to most previous
observations.  \cyg showed root mean square (rms) variability $\approx
30\%$ characterized by a power spectral density (PSD) that was nearly flat
between $0.02$--$0.2$\,Hz.  The PSD was approximately $\propto f^{-1}$
between $\approx 0.2$--$2$\,Hz, while the power law index of the
$2$--$90$\,Hz PSD was seen to increase from $\approx -1.7$ to $\approx
-1.4$ between the lowest and highest energy bands.  The coherence function,
rarely presented for most previous observations (although see
\cite{vaughan:91a,vaughan:97a,cui:97b}), was remarkably close to unity over
a wide range of frequencies.  We also considered the Fourier
frequency-dependent time delay between soft and hard photons.

As discussed in paper II, hard photons in \cyg are seen to lag behind soft
photons with a time delay that is dependent upon Fourier frequency (see
also \cite{miyamoto:89a,miyamoto:92a,cui:97b,crary:98a}).  Comparing the
($0$--$3.9$\,keV) band to the ($14.1$--$45$\,keV) band, the time delays are
approximately $\propto f^{-0.7}$ and range from $\approx 2\times
10^{-3}$--$0.05$\,s.  A more detailed study of these time delays will be
the focus of this work.  Specifically, we wish to understand what leads to
this factor of $20$ range in timescales, and we consider models where the
time lags are: created within the outer accretion disk
(\S\ref{sec:compcon}); created via photon diffusion in an extremely large
corona (\S\ref{sec:kht}); or are due to wave propagation (\S\ref{sec:prop}).

As the basis for our discussions, we will for the most part take as our
straw man model the `sphere+disk' Comptonization model that we considered
in paper I. We illustrate the geometry of this model in Figure~1.
This model has a uniform coronal heating rate, nearly uniform density
structure, but a non-uniform temperature structure.  The seed photons for
Comptonization are due to (reprocessed and direct) soft radiation from the
outer, thin disk.  This model does not include dynamical information.  The
Advection Dominated Accretion Flow (ADAF) models have a similar geometry,
as shown in Figure~1,
but they model the accretion dynamics as well
(\cite{abramowicz:95a,narayan:94a,narayan:96e,esin:97c}).  For ADAFs, a
substantial portion of the accretion energy is advected into the black
hole, potentially at a large fraction of free-fall speeds, in the form of
thermal energy.  (The density profile of matter in free-fall is $\propto
R^{-3/2}$.) The advective inner regions also produce cyclo/synchrotron
photons, which are assumed to be the seed photons for Comptonization.
Another model that we shall discuss is the Comptonization model of
\citey{kazanas:97a}, hereafter referred to as KHT (see also
\cite{boetcher:98a}).  This model postulates a large spherical corona with
a radial density profile typically $\propto R^{-3/2} \rightarrow R^{-1}$,
but uniform temperature structure.  The seed photons for Comptonization are
isotropic and can originate either within the central regions of the corona
(KHT), or externally (\cite{boetcher:98a}). Note that these models do not
consider detailed flow dynamics.

\begin{figure*}
\centerline{
\psfig{figure=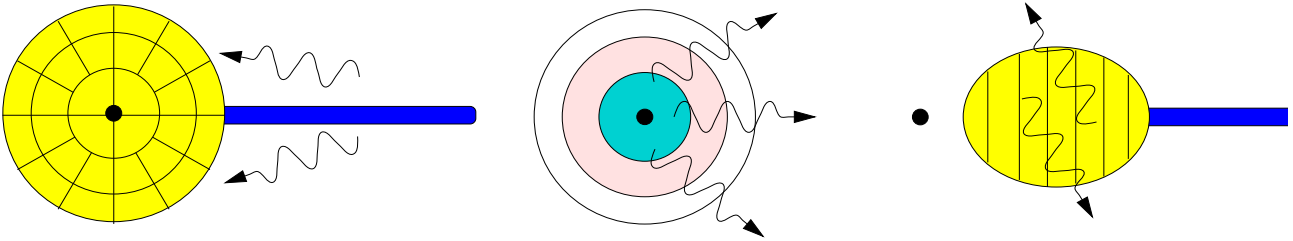,width=0.96\textwidth}
}
\caption{\small Geometries for recent models of \cyg.  {\it Left:} Sphere+disk
  geometry, as considered by Dove et al. (1998) (paper I).  The seed
  photons for Comptonization come from the outer disk. The central corona
  has a uniform heating rate and seed electron density but a non-uniform
  temperature structure.  (cf. Dove, Wilms, \& Begelman 1997).  {\it
    Middle:} Geometry considered by Kazanas, Hua, \& Titarchuk (1997).  The
  seed photons for Comptonization are isotropic and arise from an
  unspecified source within the central region of the system. The corona
  has a spherical structure with a typically $R^{-3/2} \rightarrow R^{-1}$
  density profile, but uniform temperature structure. {\it Right:}
  Advection Dominated Accretion Flow (ADAF) geometry (cf. Narayan 1996).
  The seed photons for Comptonization are cyclo/synchrotron photons that
  originate within the advective inner region.  A cylindrical symmetry
  approximation for the structure is often taken.}
\label{fig:models}
\end{figure*}

In order to understand how these models might lead to the observed time
delays, let us consider some of their characteristic timescales.  We shall
take the fiducial radius of the corona (or inner, advective region) to be
$R = 50~GM/c^2$, with $M=10~M_\odot$. (We shall adopt $M=10~M_\odot$
throughout the rest of this paper.)  We then have the following
characteristic timescales.  The radial light-crossing timescale, $t_{\rm
LC}$, is $2.5\times10^{-3}$\,s, which is comparable to the shortest time
lags.  For an 87\,keV Compton corona the radial sound crossing time,
$t_{\rm SC} = 5 \times 10^{-3}$\,s.  The free-fall timescale, $t_{\rm
FF}$, relevant for the ADAF models, is $8.5\times10^{-3}$\,s and is
proportional to $R^{3/2}$.  All three of these timescales are mainly
relevant for the shortest time delays.  In order for them to be applicable
to the longest time delays, one needs to consider a radius of $10^3~GM/c^2$
for the light-crossing timescale and a radius of $\approx 160~GM/c^2$ for
the free-fall timescale.  (Some ADAF models do posit a large radius
for the advective region; \cite{narayan:96e,esin:97c}.)  A single one of these
mechanisms can explain the observed dynamic range in time delays only if it
samples a comparably large range in radii.

The viscous diffusion timescale is $\approx R^2/\nu$, where $\nu$ is the
kinematic viscosity.  Taking an $\alpha$-disk model, $\nu \approx \alpha H
c_s$, where $H$ is the disk thickness, and $c_s$ is the speed of sound
(cf. \cite{accretion}).  Using the speed of sound from our best fit corona
model and taking $H\approx R$ and $\alpha \approx 0.1$, the viscous
diffusion time, $t_{\rm D}$, is approximately $0.03$\,s, which is
comparable to the longest time lags.  The Keplerian period at $R=50~GM/c^2$
is $t_{\rm K} \approx 0.1$\,s, and is also comparable to our longest time
delays.  All of the above characteristic timescales, along with the
observed time lags, are presented in Figure~2.

We see that there are a range of characteristic timescales comparable to
both the shortest and longest observed time lags.  The latter are perhaps
the more difficult to understand theoretically.  Two extremes for
explaining the longest time lags are: a small radius and a slow mechanism
(e.g., viscous diffusion); or a fairly large radius and a fast mechanism
(e.g., light or sound speed propagation).  The fact that there is extremely
high coherence between the soft and hard bands (cf. paper II), even at
frequencies near $0.1$\,Hz where we see the longest time delays, makes it
unlikely that a combination of these two possibilities is at work.  (Even
individually coherent processes will appear incoherent when summed, unless
each process yields the same transfer function from soft to hard;
\cite{vaughan:97a}).

\begin{figure*}
\centerline{
\psfig{figure=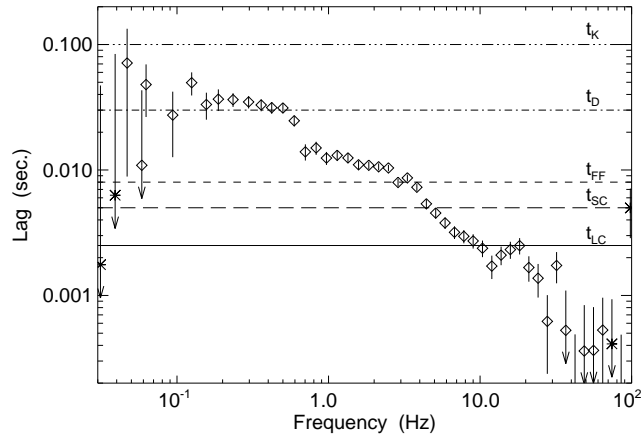,width=0.48\textwidth}
}
\caption{\small RXTE observations of time delays between the ($0$--$3.9$\,keV) and
  ($14.1$--$45$\,keV) lightcurves of Cyg~X--1.  Diamonds indicate where the
  hard lags the soft, and stars indicate where the soft lags the hard
  (paper II).  Lines represent characteristic timescales for a coronal
  radius of $50~GM/c^2$ with $M=10~M_\odot$. Solid line is the radial light
  crossing timescale ($t_{\rm LC}$); long dashed line is the radial sound
  crossing time ($t_{\rm SC}$); short dashed line is the free-fall
  timescale ($t_{\rm FF}$); dash dot line the viscous diffusion timescale
  for a hot, geometrically thick cloud ($t_{\rm D}$); and the dash triple
  dot line is the Keplerian orbital period for a thin disk ($t_{\rm K}$).}
\label{fig:chartime}
\end{figure*}

In this paper, we consider both of the above possibilities.  In
section~\ref{sec:compcon} we calculate the effect that coronal
`reprocessing' has on time delays that are intrinsic to the soft X-ray seed
photons.  We explore this possibility, first discussed by
\citey{miller:95a} and \citey{nowak:96a}, using our best-fit coronal model
of paper I.  In section~\ref{sec:kht}, we consider the suggestion of
KHT that the time delays are created by Compton scattering
in a corona that extends several decades in radius.  We present simple,
phenomenological propagation models in section~\ref{sec:prop}. We use these
models to show that the time delays can have a complicated frequency
dependence even under very simple assumptions.  We furthermore discuss the
characteristic propagation speeds that such models imply. We present our
conclusions in section~\ref{sec:sum}.

\section{Time Delays Intrinsic to the Seed Photons}\label{sec:compcon} 

As discussed in paper II (and references therein), one naturally expects
that the hard photons lag the soft photons if the high energy spectrum is
mainly due to Comptonization.  The time delay is due to the fact that the
hard photons undergo several more scattering events than the soft photons.
Such time delays should approximately depend upon the logarithm of the
energy (paper II, and references therein), and should be of order the light
crossing time of the corona.  As was first pointed out by
\citey{miyamoto:89a}, and shown in Figure~2
above, this is considerably shorter than the longest observed delays.

This fact led \citey{miller:95a} to suggest that the time lags might be
intrinsic to the seed photons for Comptonization, and that the input time
delay's frequency dependence (although not amplitude) might be preserved by
Comptonization.  As discussed by \citey{miller:95a} and elaborated upon by
\citey{nowak:96a}, the Fourier frequency-dependence of the time delay is
preserved, typically at low frequency, if the difference between the input
and output photon energies is not too great.  The amplitude of the time
delay, however, tends to be \emph{decreased} in the scattering process,
perhaps substantially so (\cite{nowak:96a}).  A constant time delay is
introduced typically at high Fourier frequency. The amplitude of this time
lag `shelf' is given by the difference of the mean diffusion times through
the Compton cloud for the two energy bands being compared, and thus should
depend logarithmically upon energy (\cite{pozd:83a,miller:95a,nowak:96a}).
In addition to the introduction of a constant time delay at high Fourier
frequency, one also expects that the intrinsic PSD of the seed photons will
be attenuated, especially at high frequency
(\cite{brainerd:87a,kylafis:87a,wijers:87a,stollman:87a,bussard:88a,kylafis:89a,miller:92a,miller:95a,nowak:96a}).

Given a source of seed photons and a Comptonization model, one can
calculate these effects (\cite{miller:95a,nowak:96a}).  Take a discretely
sampled seed photon lightcurve, $s_i^{[k]}$, where $i$ denotes the time bin
and $[k]$ denotes the energy band.  The output lightcurve, $h_j^{[l]}$, is
measured at times $j$ and in energy band $[l]$.  The two can be related by
an equation of the form
\begin{equation}
h_j^{[l]} ~=~ \sum_k r_{ji}^{[lk]} s_i^{[k]} ~~,
\label{eq:tspace}
\end{equation}
where $r_{ji}^{[lk]}$ describes the {\it linear} transfer properties of the
Comptonizing medium in both time ($i \rightarrow j$) and energy
($[k]\rightarrow [l]$) (cf. \cite{nowak:96a}, and the Appendix below). The
Compton corona can be described by such a linear transfer function if the
coronal structure is \emph{stationary}, which would also imply a unity
coherence function measured between the output bands (\cite{nowak:96a}).

In Fourier space, eq.~(\ref{eq:tspace}) can be written as
\begin{equation}
H_m^{[l]} ~=~ \sum_k R_m^{[lk]} S_m^{[k]} ~~,
\label{eq:fspace}
\end{equation}
where capital letters denote discrete Fourier transforms, $m$ denotes the
discrete frequency bin, and $R_m^{[lk]}$ is the transform of a
\emph{column} of $r_{ji}^{[lk]}$ (i.e., $i$ held fixed and $j$ free to
vary; \cite{nowak:96a}).  We can then relate the measured PSD and cross
power spectrum (CPD) to the input PSD and CPD by
\begin{eqnarray}
{H_m^{[l]}}^\ast H_m^{[n]} &=& \sum_k {R_m^{[lk]}}^\ast R_m^{[nk]} 
     \left ( S_m^{[k]} \right )^2 \nonumber \\
& & + \sum_{j \ne k} {R_m^{[lk]}}^\ast R_m^{[nj]}  
     \left ( {S_m^{[k]}}^\ast S_m^{[j]}  \right )  ~~.
\label{eq:how}
\end{eqnarray}

Given a Comptonization model we can readily calculate $r_{ji}^{[lk]}$,
which can then easily be Fourier transformed to yield $R_m^{[lk]}$.  As
described in the Appendix, we have calculated the transfer functions for
our best fit Comptonization model of paper I.  We use these transfer
functions to assess the effect of our Compton corona model on an input
white noise source with an {\it intrinsic} time delay between soft and hard
photons that is $\propto f^{-1}$.  In our model of paper I, $25\%$ of the
seed photons come from energies $< 40.4$\,eV, $50\%$ come from energies $<
115$\,eV, and $75\%$ come from energies $< 214.4$\,eV.  We choose these
three energies as seed photon energies, and furthermore we impose a
constant Fourier phase lag of $\pi/2$ between the $214.4$\,eV and $115$\,eV
variability, as well as between the $115$\,eV and $40.4$\,eV variability.
The resulting phase lag of $\pi$ between the $214.4$\,eV and $40.4$\,eV
variability is the {\it maximum} allowed phase lag\footnote{By convention,
the Fourier phase lags are taken to be between $[-\pi,\pi]$.  Thus a time
delay that leads to a phase lag of $3\pi/2$ will be measured as a phase
\emph{lead} of $\pi/2$.  We choose the maximum phase lag of $\pi$ in order
to determine the maximum possible output phase lag.} between any two energy
bands (see \cite{nowak:96a}, and paper II). We also choose the amplitude of
the variability to be identical for all three seed photon energy bands.

In Figure~3
we present the result of passing such white noise variability through our
Comptonization model of paper I.  As an example, we present theoretical
time delays and PSD for the ($2$--$4$\,keV) band compared to the
($12$--$50$\,keV) band.  (Neutral hydrogen column density is not included
in these theoretical calculations of time delays.)  We hold all parameters
of the Comptonization model fixed, including the temperature of the seed
photons from the disk, to the values of paper I; however, we vary the
physical radius of the corona.  Specifically, we consider radii of
$30~GM/c^2$, $50~GM/c^2$, $150~GM/c^2$, and $500~GM/c^2$.  A number of
results are immediately apparent from these figures.

First, the largest physically allowed phase lag between the $214.4$\,eV and
$40.4$\,eV variability was still \emph{too small} to reproduce the majority
of the observed time delays.  Thus the hypothesis of \citey{miller:95a} and
\citey{nowak:96a} that the time delays could be intrinsic to the disk
appears to be wrong.  As discussed by \citey{nowak:96a}, such a large input
phase lag is required because of the great disparity between the input
energy ($\approx 150$\,eV) and the output energies ($\approx 3$\,keV,
$20$\,keV)\footnote{Imagine that we have input energy bands, $E_a$ and
$E_b$, \emph{each} of which scatter into two observed output bands, $E_1$
and $E_2$. If the input energy bands scattered into the output energy bands
in equal proportion (i.e., $[E_a \rightarrow E_1]/[E_a \rightarrow E_2] =
[E_b \rightarrow E_1]/[E_b \rightarrow E_2]$) then the \emph{intrinsic}
time delays would be completely wiped out.  In reality the scattering is
slightly asymmetric, which allows a remnant of an input time delay to
remain.  If $E_a < E_b$ and $E_1 < E_2$, a slightly smaller proportion of
$E_a$ scatters into $E_2$, as compared to $E_b$ that scatters into $E_2$.
For fixed input energy bands, this asymmetry decreases with increasing
output band energy, and hence the intrinsic time delays are more completely
erased (cf. \cite{nowak:96a}).}.
The required input phase lag can be decreased if the temperature of the
seed photons is increased; however, this is not allowed by the energy
spectral modeling (paper I).  Further argument against the hypothesis is
garnered from the fact that the calculated time delays \emph{do not} depend
logarithmically upon energy, contrary to the observations (paper II, and
references therein).  The theoretical model does show a logarithmic energy
dependence at high Fourier frequency, where a shelf in the theoretical time
delay is clearly seen; however, the energy dependence weakens for lower
frequencies.  As shown in paper II (Figure~12), the data show a logarithmic
energy dependence even at $0.3$\,Hz.  We are thus forced to the conclusion
that the majority of the observed time delays, in one fashion or another,
must be created within the corona if our basic model for the energy
spectrum is correct. However, this is \emph{not} the same as saying that
the variability (PSD) must be directly created in the corona
(cf. \S\ref{sec:kht}).

We do expect one effect of Comptonization to remain, even if the time lags
are created within the corona itself, namely the time lag shelf at high
frequency.  After one scatter, a photon has essentially lost all
information as to the spatial location of its origin. Thus whether the seed
photons come from an outer disk or whether they are internal to the corona
(such as for ADAF models that invoke cyclo/synchrotron seed photons;
\cite{narayan:96e,esin:97c}), the difference in diffusion times to reach
two different output energies will still lead to a time lag shelf.  The
shortest observed time lag should be no smaller than this theoretically
expected shelf.  Such shelves clearly are seen at high Fourier frequency in
the theoretical models presented in Figure~3.
They range from $2 \times 10^{-3}$\,s for $R=30~GM/c^2$ to $0.04$\,s for
$R=500~GM/c^2$.  The observational data, on the other hand, do show a
flattening in the time delays in the $10$--$30$\,Hz range.  If we take this
as the upper limit to an allowed theoretical time delay, then the maximum
allowed coronal radius for our model of paper I is $\approx 30~GM/c^2$.
There is also an observed flattening of the time delay in the region of
$0.7$--$3$\,Hz.  If we take this as the upper limit to an allowed
theoretical time delay, then the maximum allowed coronal radius is $\approx
150~GM/c^2$.

Which of these observational limits should we choose, and why do we ignore
the even shorter time delays above $30$\,Hz?  As discussed in paper II, the
time delays above $30$\,Hz are especially subject to noise as both the PSD
\emph{and} the coherence function are decreasing in this regime.  We again
note that there is a `hardening' of the $2$--$90$\,Hz PSD with increasing
energy (paper II), coincident with this coherence loss.  We postulated that
this was indicative of additional, multiple variability components that
were being created directly within the corona.  One speculation would be
that these timescales are probing flares that are `feeding' the corona on
dynamical timescales.  Without being able to identify the physical nature
of these hypothesized extra variability components, we do not know what
their intrinsic time delays are.  We \emph{do} know, however, that multiple
uncorrelated variability components will lead to a loss of coherence, and
that the \emph{net} observed time delay in such regions will be a
combination of many intrinsic lags and possibly even leads
(\cite{vaughan:97a}; paper II, \S3.1).  Thus an \emph{incoherent} frequency
range \emph{can} have observed time delays \emph{less} than the minimum
theoretically expected time lag shelf.

We should then choose the maximum allowed theoretical time lag shelf to be
the \emph{minimum} time delay observed in a region of near unity coherence.
For our observations of Cyg X--1, this would be at $\approx 10$\,Hz, and
thus this limits us to a maximum coronal radius of $\approx
30~GM/c^2$. This is consistent with the limits on the PSD as well
(cf. Figure~3).
Such a small corona has little effect on the PSD in the $0.02$--$2$\,Hz
regime, where PSD for all five observed energy bands (paper II) have
roughly the same shape.  We note, however, that based upon the PSD alone,
especially if the PSD above $\aproxgt 10$\,Hz is contaminated by other
sources of variability, a much larger coronal radius ($\approx 150~GM/c^2$)
is tolerated.

It is tempting to associate the various `flattened' regions ($\approx
0.1$--$0.5$\,Hz, $\approx 0.7$--$3$\,Hz, and $\approx 10$--$30$\,Hz) seen
in the time delays of Figure~3
with time lags shelves due to Comptonization. This would imply a Compton
corona with characteristic radii of $\approx 30~GM/c^2$, $\approx
150~GM/c^2$, \emph{and} $\approx 500~GM/c^2$.  The (nearly) uniform density
coronal model of paper I {\it cannot} produce such a range of time lags.
However, recently KHT have proposed a coronal model with an
$R^{-1}$ density profile that {\it can} produce a broad dynamic range of
time delays.  We consider such models in the next section.

\begin{figure*}
\centerline{
\psfig{figure=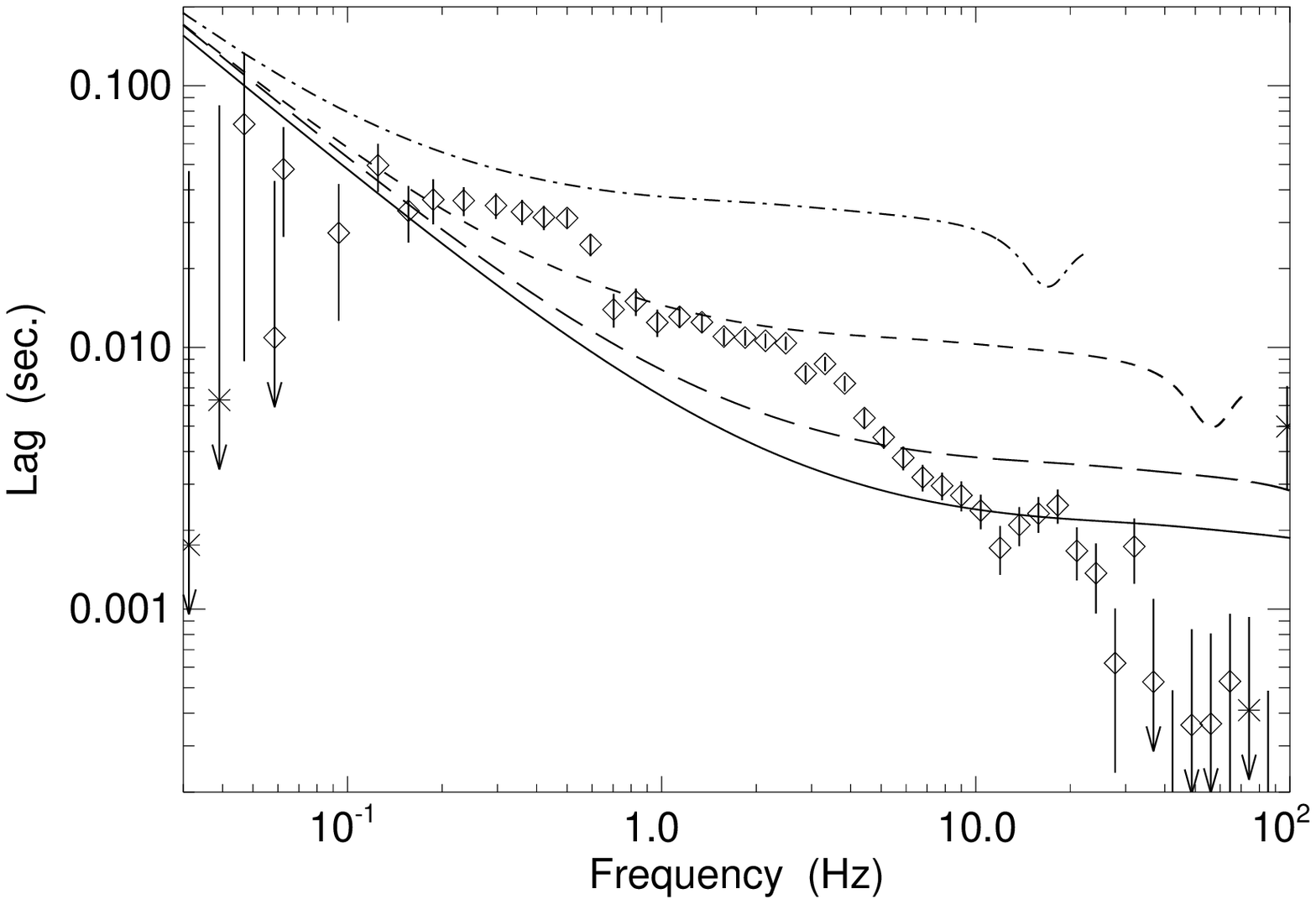,width=0.48\textwidth}
\psfig{figure=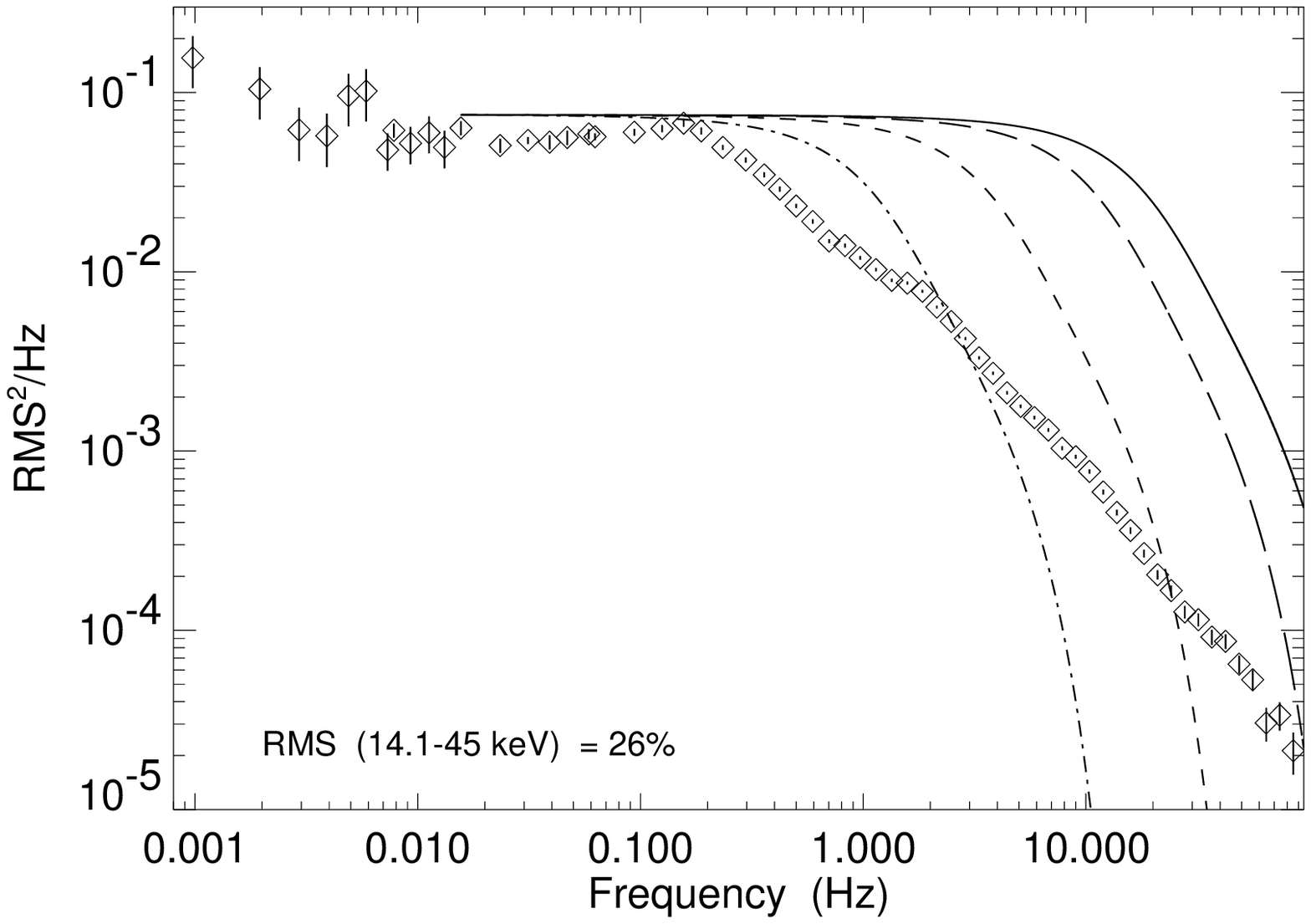,width=0.48\textwidth}
}
\caption{\small {\it Left:} Diamonds and stars are the same time lags as
  shown in Figure~2. 
  Solid lines are the theoretical time delays between the ($12$--$50$\,keV)
  and ($2$--$4$\,keV) lightcurves, assuming that the seed photons for
  Comptonization have an intrinsic phase lag of $\pi$ between the 40.4 eV
  and 214.4 eV lightcurves. (The intrinsic time lag in the seed photons is
  therefore $\propto f^{-1}$; see text.) {\it Right:} Diamonds are the
  observed Cyg~X--1 PSD in the ($14.1$--$45$\,keV) band, compared to
  theoretical ($12$--$50$\,keV) PSD for a white noise input (arbitrary
  normalization). For both figures, the solid line represents a coronal
  radius of $R=30~GM/c^2$, the long dashed line represents $R=50~GM/c^2$,
  the short dashed line represents $R=150~GM/c^2$, and the dash dot line
  represents $R=500~GM/c^2$.}
\label{fig:response}
\end{figure*}

\section{Time Delays Created by the Corona}\label{sec:kht} 

In the model of KHT, as well as the models of \citey{boetcher:98a}, both the
observed power spectral densities and time lags are the result of passing
an \emph{isotropically emitted} white noise (i.e. flat) spectrum through an
extended Comptonizing medium with a power-law density profile.  The case of
$\rho \propto R^{-1}$ represents equal optical depth per decade of radius,
and therefore roughly equal probability of seed photons from an isotropic
source scattering within any given radial decade.  This leads to a
power-law shape for the observed PSD, as opposed to the fairly sharp cutoff
seen in Figure~3.
Time delays are created by the difference in diffusion times through the
corona for hard and soft photons.  Photons that scatter over large radii
will have their intrinsic high frequency variability wiped out (as in
Figure~3);
therefore, any observed high-frequency variability must be due to photons
that scattered only within the inner radial regions. High frequency
variability thus exhibits short time delays between hard and soft photons.
Low frequency variability potentially can be observed from photons that
have scattered over large radius.  Low frequency variability thus exhibits
longer time delays between hard and soft photons.  The time delay at all
Fourier frequencies is expected to increase as the logarithm of the ratio
of the hard to soft energy, as is observed (cf. paper II, and references
therein).  For quantitative agreement with the observations, KHT require
coronal radii of ${\cal O}(10^4~GM/c^2)$.

The main objection to this scenario is that the source of soft seed photons
is not fully specified.  ASCA observations of the hard/low states of \cyg
(\cite{ebisawa:96b}) and of GX339--4 (Wilms et al. 1998, in preparation)
show evidence of a `soft excess' that is reasonably well-modeled by a
multicolor blackbody spectrum with a peak temperature of $\sim 150$\,eV.
This temperature is suggestive of--- but not definitive evidence for--- the
inner edge of an accretion disk (the putative source of the seed photons,
although see \cite{narayan:96e,esin:97c}) being at radii $\aproxgt
50~GM/c^2$, as opposed to being at the center of the corona.  Both of these
sources, as well as several other low/hard state GBHC have shown evidence
of \emph{weak} and \emph{narrow} Fe lines at 6.4\,keV (\cite{ebisawa:96b};
Wilms et al. 1998, in preparation).  The properties of these
lines\footnote{Note that if the energy release from the corona is centrally
  concentrated, as in the ADAF models or the coronal model with $R^{-1}$
  heating described below, then the Fe line places \emph{upper limits} to
  the coronal radius (cf. \cite{esin:97c}).  The fact that Fe lines are
  seen with equivalent widths $\approx 30$--$40$\,eV, even in very low
  luminosity observations of GBHC such as Nova Muscae (\cite{zycki:98a})
  and GX339--4 (Wilms et al. 1998, in preparation), tends to argue against
  very large radii in the sphere+disk geometry.} can be explained naturally
in the sphere+disk geometry.

\begin{figure*}
\centerline{
\psfig{figure=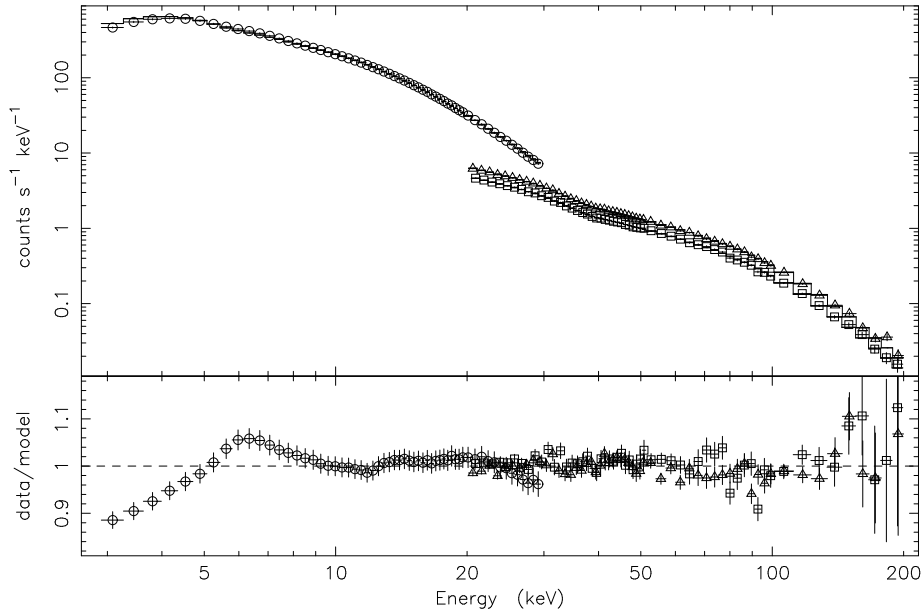,width=0.75\textwidth,angle=270}
}
\caption{\small Best fit `sphere+disk' Comptonization model to RXTE data of
  Cyg~X-1.  (Circles are PCA data, triangles and squares are HEXTE data;
  cf.  paper I.)  Unlike the Comptonization models presented in paper I,
  here the heating rate per particle is $\propto R^{-1}$ and the coronal
  density structure is $\propto R^{-1}$.  The best fit parameters were a
  total optical depth of 2.9 and an average coronal temperature of 58\,keV.
  The reduced $\chi^2$ for the fit was 1.75.  As discussed in paper I, a
  possible reason for the residuals in the 3--7\,keV region (which are
  slightly larger than the systematic uncertainties in the PCA) is our
  assumption of a sharp transition from thin, cold disk to thick, hot
  corona.}
\label{fig:khtspec}
\end{figure*}

Does a corona with a power law density profile, but with seed photons
arising from a geometrically thin outer disk, still produce the
characteristic PSD and time delays as described by KHT?  To answer this
question, we have created a grid of Comptonization spectra with the
following properties.  The heating per particle in the corona was taken to
go as $R^{-1}$ (i.e., proportional to gravitational energy release), the
density was taken to go as $R^{-1}$ (i.e., the density structure chosen by
KHT), and the seed photons were taken to come from a multicolor blackbody
spectrum, with peak temperature 150\,eV, originating from a geometrically
thin outer disk.  We used the same Comptonization code as described in
\citey{dove:97a} and used in paper I.  Unlike the models of KHT, which were
isothermal out ${\cal O}(10^4~GM/c^2)$, our simulations include the
derivation of the self consistent temperature structure of the corona by
balancing the local Compton cooling rate with the local heating rate (which
was $\propto R^{-1}$). For typical parameter values, the temperature was
about 72\,keV above the pole of the corona, and only 40\,keV at the equator
near the edge of the sphere. (This cooler temperature is due to the
increased density of soft photons near the accretion disk.) We applied
these models to our RXTE spectra of \cyg (cf.  paper I), and obtained a
reasonable fit to the data (Fig.~4).
We find a best fit coronal optical depth of 2.9, with an average coronal
temperature of 58\,keV.

As described in the Appendix, we again used a linear Monte Carlo code (with
the radial coronal structure taken from our best fit nonlinear model, but
excluding the pole to equator temperature gradient) to calculate the
time-dependent transfer functions for seed photon energies to transit to
observed hard X-ray energies.  In an analogous manner to the calculation
described in \S\ref{sec:compcon}, we used these transfer functions to
determine the effect that the corona has on any variability inherent to the
Comptonization seed photons.  In Figure 5a,
we show the resulting time lags between soft and hard X-ray variability. In
Figure 5b,
we show the resulting PSD assuming a white noise power spectrum for the
variability of the seed photons.  Here we take all the photons, an input
blackbody with temperature $kT=150$\,eV, to have a uniform initial Fourier
phase. In both of these figures, we present results for coronal radii
ranging from 30--$500~GM/c^2$.

Note that unlike the models of KHT and \citey{boetcher:98a}, we do
\emph{not} find time lags that are proportional to Fourier period and we do
\emph{not} find a PSD with a power law dependence over a wide range of
frequencies.  Our results are qualitatively and quantitatively similar to
the uniform (in density and heating) coronal models presented in Figure~3.
Specifically, the resulting time lag is nearly constant as a function of
Fourier frequency, and is comparable to the light crossing time across the
diameter of the entire corona.  The resulting PSD (assuming a white noise
input) has a sharp cut-off as opposed to a power law shape.

The major difference between the model presented here and those of KHT and
\citey{boetcher:98a} is one of geometry.  The latter models assume an
isotropic source of seed photons. The models of KHT assume that the seed
photons originate from interior to the corona, while the models of
\citey{boetcher:98a} consider both internal and external illumination of
the corona.  Both models, assume isotropic illumination.  The core of the
corona, where the photons can undergo the scatters on the shortest
timescales and thus not suffer substantial losses of high frequency
variability power, is initially visible to the seed photons in these
models.  

In our model, where the seed photons originate in a geometrically thin disk
exterior to the corona, the central core of the corona subtends a very
small solid angle as viewed by the disk.  The photons emanating from the
disk do not isotropically illuminate the corona.  This geometry, coupled
with the fact that our best-fit model is mildly optically thick, means that
a substantial fraction of the photons must first scatter on the large radii
of the outer corona before being able to scatter within the inner radii of
the corona.  Thus the time lags are dominated by the \emph{longest} time
delays at \emph{all} Fourier frequencies.

Further differences with our model are that we consider coronal heating
$\propto R^{-1}$, we allow reprocessing of hard X-ray photons in the
geometrically thin, outer accretion disk, and we allow for anisotropic
temperatures in the corona (although in the linear Monte Carlo code we only
include the radial temperature gradients).  We conjecture that the primary
reason the results of KHT (who use a uniform coronal temperature of
$50$\,keV and optical depth of $\tau=2$) are so drastically different is
precisely because of their choice of an isotropic source of seed photons.
[\citey{boetcher:98a} show that results qualitatively similar to those of
KHT are obtained even if the isotropic source of seed photons is exterior
to the corona.]  Again, the physical nature of this source is not fully
described in the work of KHT.  However, such a seed photon source geometry
is qualitatively similar to that predicted by ADAF models
(\cite{narayan:96e,esin:97c}). In our model the seed photons only
`isotropize' (as viewed by the corona) after scattering on large radii, and
therefore always exhibit the longest time delays.

\begin{figure*}
\centerline{
\psfig{figure=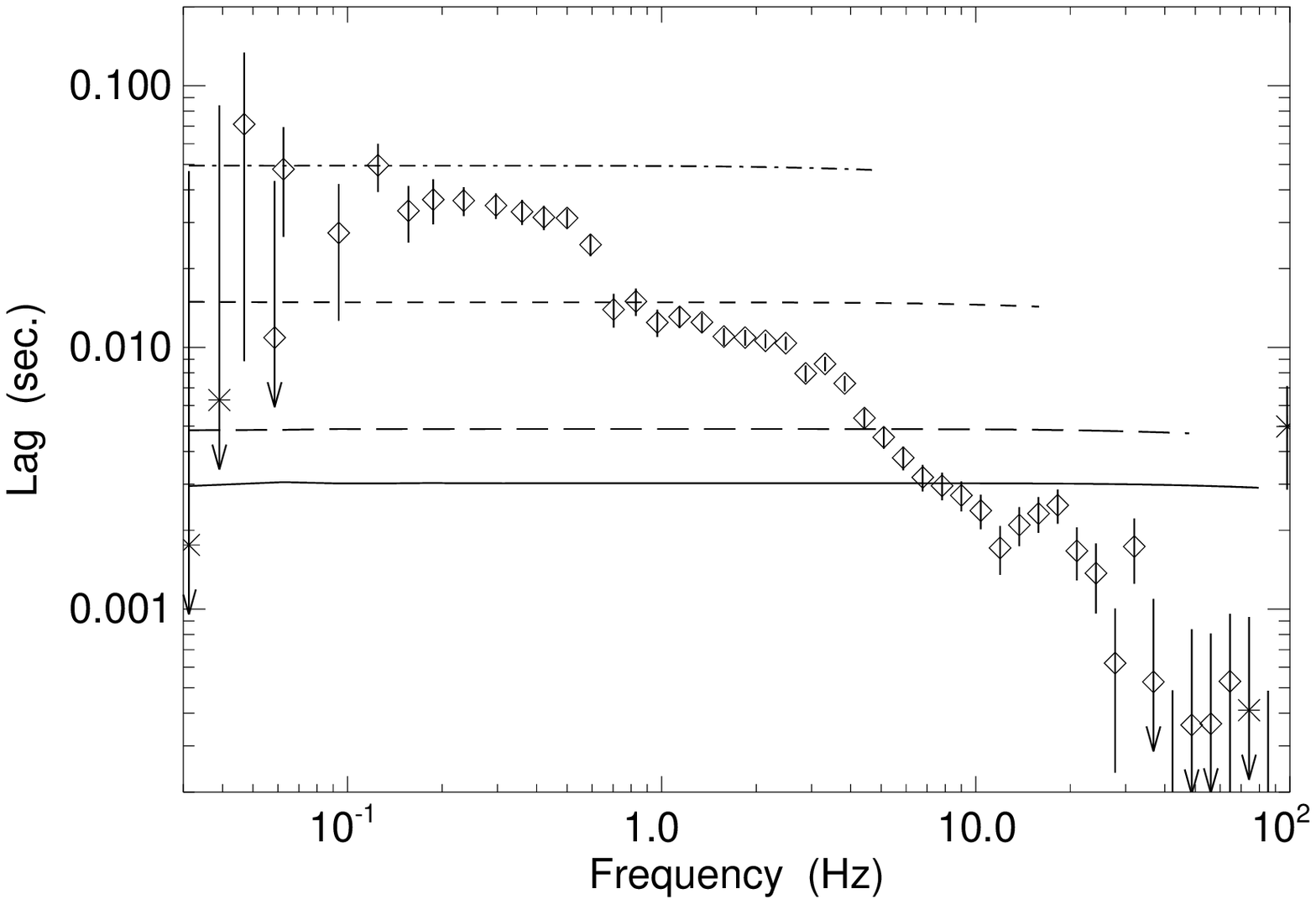,width=0.48\textwidth}
\psfig{figure=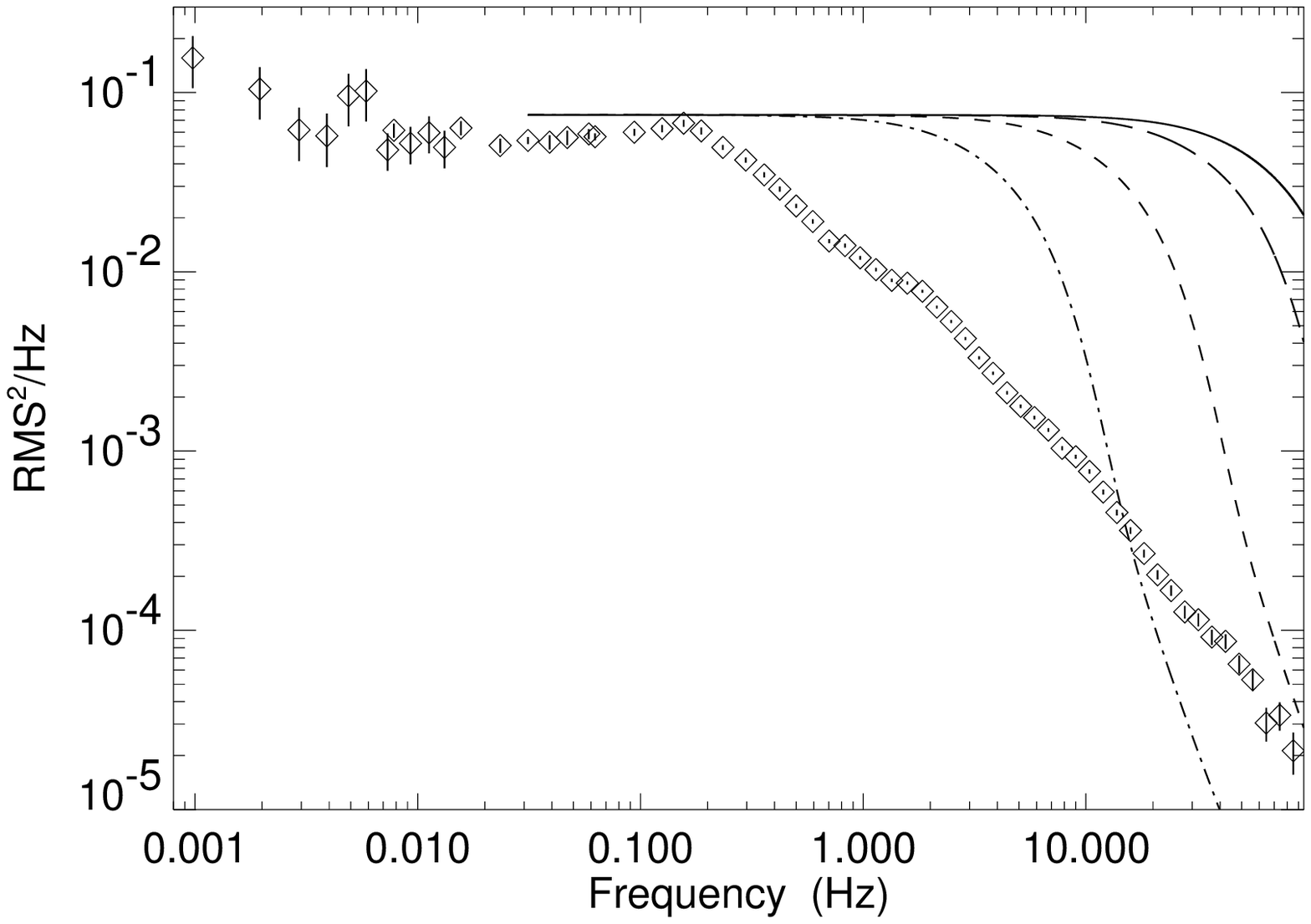,width=0.48\textwidth}
}
\caption{\small {\it Left:} Diamonds and stars represent the same time lags
  as shown in Figure~2.  
  Lines are the theoretical time delays between the ($14.1$--$45$\,keV) and
  ($1$--$3.9$\,keV) lightcurves assuming that the seed photons originate in
  the geometrically thin outer disk and have a temperature of $kT =
  150$\,eV. {\it Right:} Diamonds are the observed Cyg~X--1 PSD in the
  ($14.1$--$45$\,keV) band. Lines are the theoretical PSD for the
  ($14.1$--$45$\,keV) band assuming that the $kT = 150$\,eV seed photons
  have a white noise power spectrum (arbitrary normalization).  For both
  figures, the solid line represents a coronal radius of $R=30~GM/c^2$, the
  long dashed line represents $R=50~GM/c^2$, the short dashed line
  represents $R=150~GM/c^2$, and the dash dot line represents
  $R=500~GM/c^2$.}
\label{fig:khtresp}
\end{figure*}

\section{Propagation Models}\label{sec:prop}

It has been suggested that the observed variability in \cyg can be produced
in the context of ADAF models by disturbances propagating from the cold,
outer disk into the hot, advection-dominated inner region
(\cite{manmoto:96}).  \citey{manmoto:96} considered the inward propagation,
and advection, of a large-amplitude, cylindrically symmetric disturbance.
The basic concept of an inward propagating wave presented in
\citey{manmoto:96} is an intriguing one.  If the cold outer disk, which is
not \emph{directly} observed by RXTE, is the source of disturbances
\emph{and} the inner region then responds to these disturbance in a
\emph{linear} fashion, then we expect there to be unity coherence between
hard and soft photons (\cite{vaughan:97a}).  As also discussed by
\citey{vaughan:97a}, two-dimensional waves can reproduce some of the
qualitative features of the observed time delays.

The work of \citey{manmoto:96} was not directly applicable to {Cyg~X--1},
however, as it only considered modulation of the thermal bremsstrahlung
emission, which is essentially negligible in ADAF models; it did not
quantitatively address the lags between hard and soft X-ray photons; and
their model only produced modulation on order of the advective timescale,
and not on the broad range of timescales required to explain the \cyg PSD.
The basic concept, however, is worth considering further.  Lacking a
dynamical theory associated with our Comptonization models, we 
explore below the notion of propagating disturbances via the use of simple
phenomenological models.

We consider two-dimensional waves not only because they are naturally
suggested by a disk geometry (i.e., the hypothesized source of
disturbances), but also because even in the absence of any dispersive
mechanisms, two-dimensional waves do not satisfy Huygen's principle
(cf. \cite{morseI}).  This means that even in a nondispersive medium with
uniform propagation speed, we do not expect there to be a constant time
lag between hard and soft photons (cf. paper II, \S5.1;
\cite{vaughan:97a}).  A `cylindrical-symmetry' approximation has also been
used for the inner advective regions of ADAF models
(cf. \cite{narayan:96e,esin:97c}, and references therein).  Note for the
discussion that follows, much of the qualitative behavior that we describe
is specifically related to the two-dimensional nature of the waves.  We
therefore would not expect the behavior described below, for example, if
the waves originated in a geometrically thick outer disk.

For simplicity, we shall consider waves propagating cylindircally
symmetrically in a medium with a \emph{uniform} propagation speed.
Furthermore, we shall consider waves that propagate inward toward a
``sink'' located at the origin.  Take a disturbance, $\Psi(\vec x, t)$,
that obeys the wave equation
\begin{equation} 
\left (~ {{\partial^2}\over {\partial \vec x~  ^2}} ~-~
c_p^{-2} {{\partial^2}\over {\partial t ^2}} ~ \right ) ~ \Psi(\vec x, t)
~~=~~  -4 \pi \rho_S(\vec x, t) ~~.
\label{eq:wave}
\end{equation}
The sink at the origin is represented by $\rho_S(\vec x, t)$, which we take
to be $\equiv \delta(\vec x) \rho_S(t)$.  That is, we will
consider waves that propagate inward from the (unobserved) outer disk, and
then are absorbed at the origin \emph{without} reflection.

We can relate the disturbance, $\Psi$, to the sink, $\rho_S$, via an
advanced Green's function (\cite{morseI,vaughan:97a}).  Furthermore, let us
take the \emph{observed} soft and hard lightcurves, $s(t)$, $h(t)$, to be
the disturbance, $\Psi(\vec x, t)$, multiplied by response functions,
$g_s(\vec x)$, $g_h(\vec x)$, integrated over the disk.  Given these
assumptions, the observed soft X-ray light curve becomes:
\begin{eqnarray}
s(t) &=& (2 \pi)^{-1} \int d\vec x ~ g_s(\vec x) \Psi(\vec x, t)
        \nonumber \\
     &=& (2 \pi)^{-1} \int dt'~ \rho_{S}(t') ~~ \times
        \nonumber \\
     &&  \int dr
     df ~ r~ g_s(r) ~G_f (2 \pi f, r) ~\exp(-i 2 \pi f
     \tau) ~~,
\label{eq:lcurve}
\end{eqnarray}
and similarly for $h(t)$.  In eq.~(\ref{eq:lcurve}) we have taken the sink
to be a delta-function at the origin, we have taken the soft lightcurve
response to be cylindrically symmetric, and we have written the Green's
function, $G_f$, in Fourier space (cf. \cite{morseI}).  We do this because
we wish to calculate the time delays as a function of Fourier frequency.

Using the convolution theorem (cf. \cite{morseI}), we write $S(f)$, the
Fourier transform of $s(t)$, as
\begin{equation}
 S(f) ~=~ {{P_{S}(f)}\over{2 \pi}} ~ \int ~ d r ~ r
 ~ g_s(r) ~G_f (2 \pi f, r) ~~, 
\end{equation}
where
$P_{S}(f)$ is the Fourier transform of $\rho_{S}(t)$.  We obtain a
similar expression for $H(f)$, the Fourier transform of the hard X-ray
light curve.

The Fourier transform of $t_r(\tau)$, the transfer function between soft
and hard X-rays (cf. paper II), is then just the ratio between $H(f)$ and
$S(f)$.  That is,
\begin{equation}
T_r(f) ~=~ {{\displaystyle \int ~ d r ~r~ g_h(r) 
     ~G_f (2 \pi f, r) }\over
     {\displaystyle \int ~ d r ~r~ g_s(r) 
     ~G_f (2 \pi f, r)}} ~~.
\label{eq:trf}
\end{equation}
Note that the above does not depend upon $P_{S}(f)$.  That is, we can know
the relative amplitude and phase of $S(f)$ and $H(f)$ without actually
knowing their absolute values individually.  Furthermore, if $g_s$, $g_h$,
and $G_f$ do not vary with time, then $T_r(f)$ is constant and coherence is
preserved (\cite{vaughan:97a}; paper II).  For propagation models such as
this, the observed Fourier phase delay is simply the Fourier phase of
eq.~(\ref{eq:trf}), while the time delay is this phase divided by $2 \pi
f$.

Here we point out several caveats associated with the validity of equations
of the general form of eq.~(\ref{eq:trf}).  Linearity of the waves is
essential.  Inherent in our assumption of linearity is that the accretion
system \emph{locally} produces radiation in response to the wave, and that
the waves themselves do not produce X-rays in the outer regions of the
disk.  If the waves steepen into shocks, as for the disturbances discussed
by \citey{manmoto:96}, then a transfer function formalism will \emph{not}
be valid.  Furthermore, the coherence function for such a case would not be
close to unity, contrary to the observations (paper II;
\cite{vaughan:97a}).  Even with the assumption of linearity, we are further
relying upon the assumption of cylindrical symmetry. There cannot be too
large a variation of phase along the azimuthal direction of the wavefront,
else the inward propagating wave fronts will add incoherently.
[\citey{vaughan:97a} show that individually coherent, linear processes can
produce a net observed incoherent process when added incoherently in such a
manner. \citey{nowak:98c} discuss how such a sum of coherent processes can
be used to reproduce the small-amplitude deviations from unity coherence
seen in the lightcurve of the BHC GX~339--4.]

We shall now consider a specific example of a simple phenomenological model
for wave propagation, with parameterized responses of the soft and hard
X-rays, that qualitatively reproduces the time delays observed in {Cyg~X--1}.
This model also gives us insight into the quantitative limits that one
might be able to set on the disturbance propagation speeds from combined
dynamical/spectral models.  Taking a constant propagation speed, and
solving eq.~(\ref{eq:wave}) in terms of separate Fourier components
(cf. \cite{morseI}), the Fourier transform of the two-dimensional Green's
function becomes 
\begin{equation}
G_f(2 \pi f, r) ~=~ i \pi ~ H_0^{(2)}(kr) ~~, 
\end{equation}
where $k^2 \equiv (2 \pi f/c_p)^2$, and $H_0^{(2)}$ is the second Hankel
function of order zero. The second Hankel function has the appropriate type
of singularity at the origin, and is the relevant function for waves
traveling toward the origin (\cite{morseI}).

For illustration, let us take $g_s(r)$ and $g_h(r)$ to be given by
\begin{eqnarray} 
g_s(r) &\propto& \theta \left ( {{r - 6 r_0} \over {r_0} } \right ) *
               \exp \left ( -\alpha_s {{r}\over{r_0}} \right ) ~~,~~
        \nonumber \\
g_h(r) &\propto& \theta \left ( {{r - 6 r_0} \over {r_0} } \right ) *
               \exp \left ( -\alpha_h {{r}\over{r_0}} \right ) ~~,
\label{eq:resp}
\end{eqnarray}
where $\theta(x)$ is a step-function, and $\alpha_s$, $\alpha_h$, and $r_0
~(\equiv GM/c^2, ~M=10~M_\odot)$ are constants.  We shall consider two sets
of parameter values: ($\alpha_s = 0.06$, $\alpha_h = 0.024$), and
($\alpha_s = 0.06$, $\alpha_h = 0.048$). With these parameters, 95\% of the
soft photons come from $r \aproxlt 50~r_0$ and 95\% of the hard photons
come from $r \aproxlt 20~r_0$ and $r \aproxlt 40~r_0$, respectively.  The
resultant phase lags then depend upon $c_p$, the propagation speed of the
disturbances.

Results for this model and several different propagation speeds are
presented in Figure~6.
As can be seen from the figure, this phenomenological model qualitatively
reproduces the functional form of the observed time lags.  This is not
surprising in that the two-dimensional wave propagation Green's function
has properties similar to the `constant phase lag' transfer function
described in paper II, \S4.1 [eq. (11)]. Specifically, the constant phase
lag transfer function was seen to be a delta-function plus a $\tau^{-1}$
tail (i.e. most photons are simultaneous, with a fraction of the hard
photons lagging behind).  The two-dimensional wave Green's function in the
time domain is given by a step function, propagating at speed $c_p$,
followed by a $\tau^{-1/2}$ tail (cf. \cite{morseI}).  As we have taken the
wave to propagate cylindircally symmetricaly from the outside in, and we
have taken the soft response to extend to larger radii than the hard
response, the hard naturally lags the soft. If progressively higher energy
responses have progressively smaller radial extents, the time lags will
increase with energy.  As the energy generation in a disk goes as $R^{-1}$,
this is not an unreasonable expectation.

The most interesting things to note about Figure~6
are the propagation speeds required to quantitatively reproduce the longest
observed time delays.  We see that propagation speeds of ${\cal
  O}(1$--$10\%~c)$ are required, depending upon the degree of overlap
between the soft and hard X-ray response.  As discussed in
\S\ref{sec:intro}, this is much slower than expected for Compton
scattering, sound speed propagation, or gravitational free-fall.  The
required velocity is increased if the degree of overlap between hard and
soft response is decreased and/or if the overall radial extents of the
responses are increased.  For the former possibility, we note that we
present a model where the radial extents of the hard and soft region differ
by a factor of two and a half, and still the required propagation speed is
quite low.  For the latter possibility, it is unlikely that we can greatly
increase the overall radial extents of the responses and still have
sufficient high frequency variability in the source of the disturbances.
As discussed in paper II, the PSD has the same functional form out to
$\approx 2$\,Hz in all energy bands.  Therefore, the source of variability
(i.e., the initial inward propagating waves) must have power out to at
least this frequency.  If we consider the dynamical timescale this
corresponds to a radius of $135~GM/c^2$.  Thus it is unlikely that we can
increase the radius, and thereby increase the required propagation speed,
by more than a factor of two or three.  We note that we have tried
different functional forms for the response functions (such as step
functions) than those presented in eq.~(\ref{eq:resp}).  The results are
qualitatively and quantitatively very similar.

\section{Summary}\label{sec:sum}

In previous papers we have presented analyses of RXTE data of Cyg X--1,
where we have discussed spectral models (paper I) and timing analysis
(paper II).  For the spectral models we concentrated on `sphere+disk'
Comptonization models, as illustrated in Figure~1.
Such models were found to provide a reasonably good description of the data
over a broad energy range (3--200\,keV).  In paper II we presented power
spectral densities (PSD), the coherence function between hard and soft
variability (cf. \cite{vaughan:97a}), and the Fourier frequency-dependent
time delay between hard and soft variability. In this work we considered
this time delay in light of our `sphere+disk' Comptonization models for the
spectrum.

\begin{figure*}
\centerline{
\psfig{figure=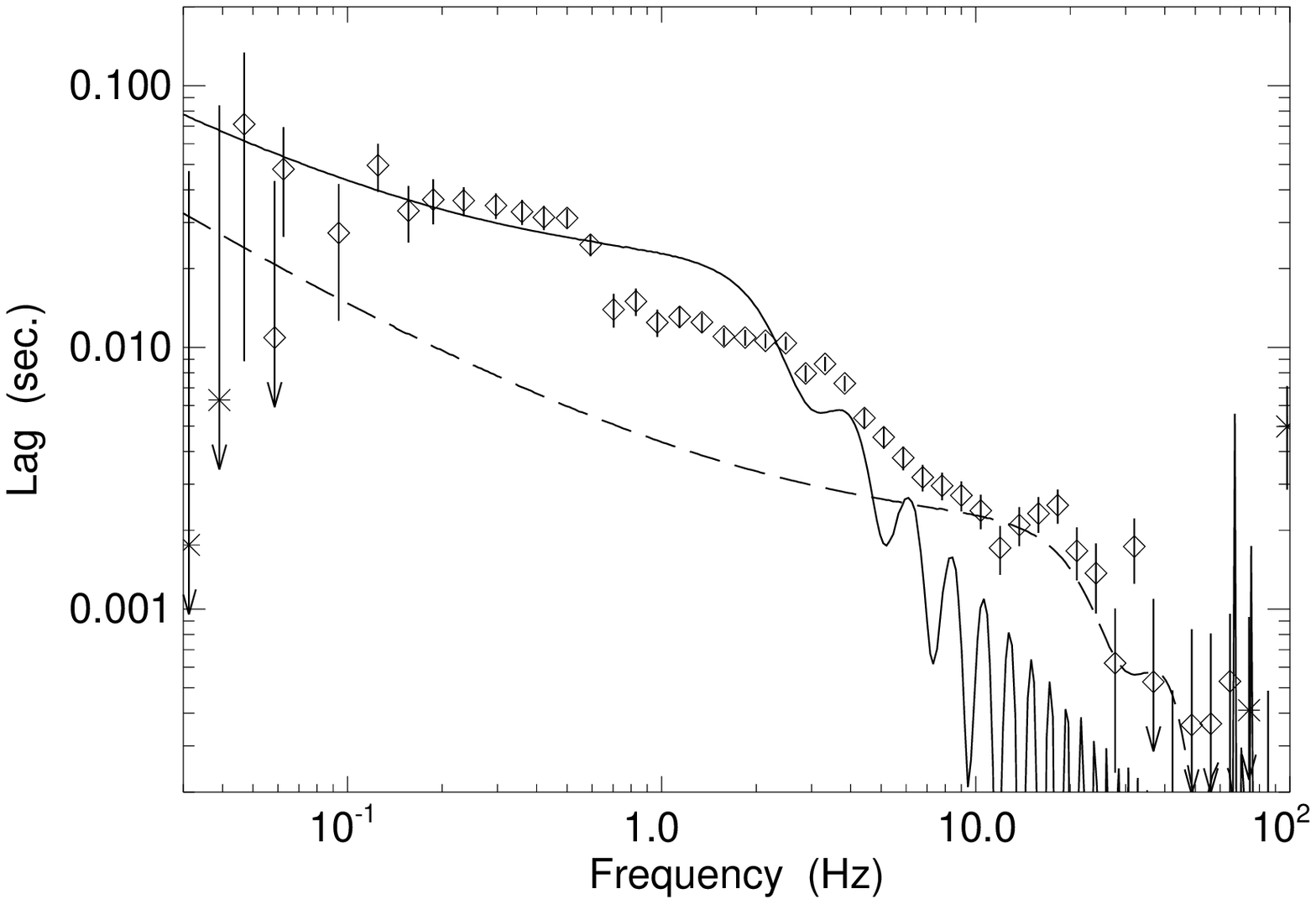,width=0.48\textwidth}
\psfig{figure=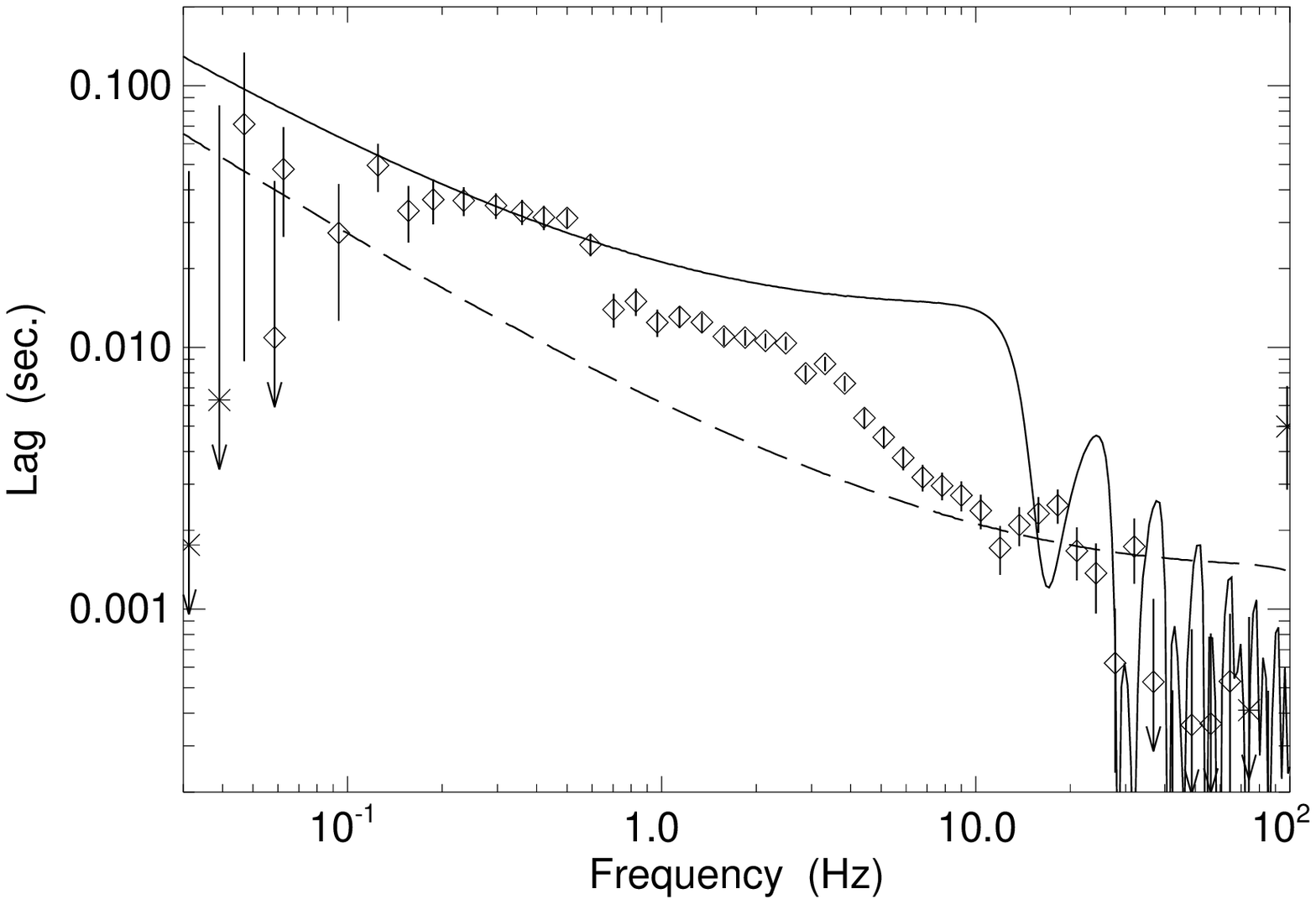,width=0.48\textwidth}
}

\caption{\small {\it Left:} Diamonds and stars represent the same time lags
  as shown in Figure~2.  
  Lines are the theoretical time delays from simple phenomenological
  propagation models.  {\it Left:} 95\% of the soft response comes from $r
  \aproxlt 50~GM/c^2$, and 95\% of the hard response comes from $r \aproxlt
  40~GM/c^2$ ($\alpha_s = 0.06$, $\alpha_h = 0.048$).  Solid line is the
  theoretical time lag for $c_p = 0.01~c$, and the dashed line is for $c_p
  = 0.1~c$.  {\it Right:} 95\% of the soft response comes from $r \aproxlt
  50~GM/c^2$, and 95\% of the hard response comes from $r \aproxlt
  20~GM/c^2$ ($\alpha_s = 0.06$, $\alpha_h = 0.024$).  Solid line is the
  theoretical time lag for $c_p = 0.06~c$, and the dashed line is for $c_p
  = 0.6~c$.}
\label{fig:prop}
\end{figure*}

The simplest expectation, as first noted by \citey{miyamoto:89a}, is that
time delays between hard and soft photons should be due to differences in
diffusion times through the Compton corona and should be nearly
\emph{independent} of Fourier frequency.  This is counter to the
observations (paper II, and references therein), which led
\citey{miller:95a} and \citey{nowak:96a} to suggest that perhaps the time
delays are intrinsic to the disk and are merely `reprocessed' by the
corona.  We explored this possibility with our Comptonization models, and
found this \emph{not} to be the case.  The required input phase lags are
unphysically large ($\aproxgt \pi$), and furthermore the resulting time
delays do not have the required logarithmic energy dependence (cf. paper
II, and references therein).  This is the chief conclusion of our paper:
\emph{if the basic sphere+disk Comptonization geometry is correct, then
the time delays must be created directly within the corona}.  

We then explored whether a corona with a power-law structure could
reproduce the observed time delays, as was first suggested by KHT.  As
opposed to this work which utilized an isotropic, central source of soft
seed photons, we again considered the sphere+disk geometry.  The seed
photons in this case comse from the geometrically thin, outer disk and
therefore are \emph{not} isotropic.  The seed photons must first scatter on
the largest radii of the corona before they can be viewed
quasi-isotropically by the inner regions of the corona.  We found that this
`sphere+disk' geometry, even with a power-law density profile, therefore
does not reproduce either the observed PSD or the observed time lags.
\emph{We conjecture that an isotropic source of seed photons, as viewed by
  the corona, is required in order to create time lags in the manner
  suggested by \citey{kazanas:97a}.}

KHT do not have a fully self-consistent model for the
source of the soft seed photons; however, their geometry is qualitatively
similar to that of the ADAF models (cf. \cite{narayan:96e,esin:97c}) which
use synchrotron photons from within the advective flow as the seed photons
for Comptonization (Figure~1).
As for the models of KHT, the ADAF models can require a very large coronal
radius.  Such a large radius, however, poses problems with interpreting the
observed spectra of GBHC.  Specifically, one typically sees a `soft excess'
with characteristic temperatures of $kT\approx150$\,keV (possible evidence
of the accretion disk), as well as weak, narrow 6.4\,keV iron lines with
equivalent widths of ${\cal O}(30\,{\rm keV})$
(\cite{ebisawa:96b,zycki:98a}; Wilms et al. 1998b, in preparation).  ADAF
models predict\footnote{There has been discussion at recent scientific
  meetings of the possibility of adding `cold blobs' of matter into the
  inner advective region of ADAF models in order to reproduce the observed
  Fe line characteristics.  Many questions are raised by such additions to
  the ADAF model, such as: what is the `natural' filling factor of the
  blobs? Will the additional cooling from the soft flux collpase the ADAF
  solution into a radiatively efficient state? Can a \emph{narrow} Fe line,
  as is suggested by the data, be produced by blobs being advected with the
  flow? Will variability associated with blobs moving over a large range of
  radii still produce near unity coherence between soft and hard radiation
  as is observed?  Considering these issues is beyond the scope of this
  current work.}  lower characteristic temperatures for any soft excess as
well as smaller equivalent widths for the iron line, if the radius of the
advective region is as large as the ${\cal O}(10^4~GM/c^2)$ required to
reproduce the longest time lags.

If we take the \emph{minimum} time delay (observed in a region where the
hard and soft variability are \emph{coherent} with each other) as
indicative of the \emph{maximum} allowed coronal radius, then \emph{the
  corona must have a small\footnote{Such a small radius is also suggested
    by the timing analysis that we presented in paper II.  Specifically,
    the slope of the the high frequency ($\aproxgt 3$\,Hz) PSD was seen to
    flatten with increasing photon energy (possibly indicative of `feeding'
    the corona on dynamical timescales at radii $\aproxlt 50~GM/c^2$) Also,
    the coherence was seen to drop at low frequency ($\aproxlt 0.02$\,Hz;
    possibly indicative of the viscous timescale of the cool, thin
    accretion disk at radii $\aproxgt 50~GM/c^2$), as well as at high
    frequency ($\aproxgt 10$\,Hz).}  radius $\aproxlt 30~GM/c^2$.}  This
led us to explore the possibility that the time delays are related to
propagation of cylindrically symmetric linear disturbances through a small
corona.  [Such `propagation models' have been considered by
\citey{manmoto:96}, for example, but see our comments in \S\ref{sec:prop}
above.] \emph{We concluded that if the corona is small and the time delays
  are due to linear disturbances propagating cylindrically symmetrically
  through the corona, then the propagation speeds are extremely slow}.
Such slow propagation speeds are likely inconsistent with ADAF models with
advective region radii $\aproxlt 150~GM/c^2$.

The advent of RXTE allows us to obtain a very broad band spectrum ($\approx
3$--$200$\,keV) and simultaneously allows us to obtain temporal data on
timescales comparable to the dynamical timescales of the very innermost
regions of GBHC systems.  The spectral capabilities of RXTE have demanded
an increasing level of sophistication from Comptonization models.
Combining the spectral data with the temporal data now requires us to
consider the \emph{dynamical} structure of Comptonization models as well.
Currently there are two broad classes of models for the observations: ADAF
models and similar coronal models with large radii (e.g.
\cite{kazanas:97a}), or sphere+disk models with fairly small radii and
(here hypothesized) relatively slow `propagation speeds' in the coronal
region.  The former models require further study to show that they agree
with \emph{all} aspects of the spectral data (and the ADAF models require
further work to demonstrate that they agree with the temporal data as
well), whereas the latter models, which are purely spectroscopic in nature
at the moment, need to be coupled with a viable dynamical theory.  Further
RXTE observations--- coupled with future observations from instruments such
as AXAF and/or XMM to study the details of the soft excesses and the weak,
narrow iron line features--- will likely be required to determine which of
the above possibilities, if any, is the most promising model for Cyg X--1
and other similar GBHC.

\acknowledgements 

We would like to acknowledge useful discussions with O. Blaes, P. Michelson,
E. Morgan, K. Pottschmidt, R. Staubert, M. van der Klis, and W. Zhang.
This work has been financed by NSF grants {AST91-20599},
{AST95-29175}, {INT95-13899}, NASA Grants {NAG5-2026}, {NAG5-3225},
{NAGS-3310}, DARA grant {50 OR 92054}, and by a travel grant to
J.W. from the DAAD.

\appendix

\centerline{A. TIME LAGS IN THE SPHERE+DISK GEOMETRY}
\bigskip

The major spectral results presented in papers~I and~II were based on
models obtained using our non-linear Monte Carlo code (cf. \cite{dove:97b}
and references therein). It is very CPU time-consuming and difficult to use
such a code to perform a study of the temporal behavior of the sphere+disk
geometry. We therefore used a separate linear Monte Carlo code for the
computation of the shots presented in this paper\footnote{In a linear Monte
  Carlo code, photons are propagated one at a time through a background
  medium with predefined properties, while a in a non-linear code, a
  multitude of particles is used to also simulate the interaction between
  the radiation field and the medium, as well as interactions between
  individual photons, such as photon-photon pair production.}. The linear
code is based on algorithms presented by \citey{marchuk:80a},
\citey{pozd:83a}, \citey{hua:86a}, and \citey{hua:97b}, Compton scattering
is simulated using the relativistic scattering formulae. The differential
Klein-Nishina cross section is used in the computation of the scattering
angle. The electrons are assumed to have a relativistic Maxwellian
distribution, the effect of which is taken into account in the simulation
of the Compton scattering and in the simulation of the mean free path of
the photon. The code also includes the interaction of the radiation with
cold matter, by making use of the fits to the photoabsorption cross
sections for the first 30 elements given by \citey{verner:95a},
\citey{verner:93a} and \citey{band:90a}. Fluorescent line emission of the
Fe~\Ka (6.4\,\keV), Fe~\Kb (7\,\keV), Si~\Ka (1.7\,\keV), and S~\Ka
(2.3\,\keV) lines was included in the simulations using the theoretical
fluorescence yields published by \citey{kaastra:93a}.

In order to reduce the statistical noise in the emerging spectrum and in
order to have a good coverage of the higher order Compton scatterings, the
method of weights was used. In this method, a photon starts out with a
weight $w=1$. After each propagation step the optical depth $\tau$ to the
boundary of the system is computed. The probability that the photon escapes
without further scatterings is $P=\exp(-\tau)$. Therefore, a photon with
weight $w\exp(-\tau)$ is added to the output-spectrum, and the rest of the
photon, with weight $w[1-\exp(-\tau)]$, continues to scatter within the
corona. After a photon gets photoabsorbed within the accretion disk, it
gets reemitted with a new weight given by $Yw$, where $Y$ is the
fluorescence yield of the absorbing element and $w$ is the weight of the
photon before the absorption, and its energy $E$ is set to the fluorescence
energy of the fluorescence line.  The photon is killed if its weight goes
below a threshold, usually taken to be $10^{-6}$.  We refer to
\citey{pozd:83a}, \citey{gorecki:84a}, and \citey{whitet:88a} for a more
in-depth description of the method of weights.

A photon emitted from the sphere can hit the disk and vice versa. These
photons are temporarily stored and dealt with after the code has finished
with following the ``primary photon''.  For a correct computation of the
spectrum emerging from the sphere+disk geometry, energy conservation of the
initial photon has to be taken into account.  Each photon escaping the
sphere and hitting the accretion disk deposits an energy $wE$ within the
disk, where $E$ is the photon energy and $w$ is again its statistical
weight. A fraction of this energy subsequently escapes the system in the
form of Compton-reflected radiation or fluorescence lines. The rest of the
deposited energy is thermalized and finally re-emitted in the form of one
or more black-body photons at the position where the original incident
photon hit the disk.

Thermalization in the accretion disk occurs primarily via photoabsorption
of a photon followed by the emission of either a photoelectron or an Auger
electron. The typical energy for these electrons is on the order of a few
keV or less. The electron then loses its energy primarily via Coulomb
interactions with other electrons. The typical relaxation timescale for
thermalization is given by (\cite{accretion}, eq.~3.32)
\begin{equation}
t  \sim \frac{\me^2 v^3}{8\pi \Ne e^4 \ln\Lambda}\,\frac{\me v^2}{2kT}
   = \frac{\me^{1/2}}{2^{3/2}\pi e^4 \ln\Lambda}\,\frac{E^{5/2}}{\Ne kT}
   \sim 10^{-5}\,{\rm s}\,
     \left(\frac{E}{10\,\keV}\right)^{5/2}
     \left(\frac{N}{10^{16}/\cm^3}\right)^{-1}
     \left(\frac{kT}{1\,\keV}\right)^{-1}
\end{equation}
where $E=\me v^2/2$ is the initial energy of the electron, $\Ne$ the
electron number density, $e$ the elementary charge, $T$ the temperature of
the plasma, and $\ln\Lambda \approx 15$ is the Coulomb logarithm. Since the
thermalization timescale is small compared to the light crossing time of the
spherical disk and the light travel time from the accretion disk to the
sphere, it can be assumed that the thermalization and Compton reflection
occur quasi-instantaneously.  Consequently the time spent by the photons
within the accretion disk was not taken into account in the determination
of the time lag and the thermalized photons were considered to be
re-emitted at the time when they hit the disk.  The effect of
thermalization is to cause ``echoes'' in the temporal response of the system
to an initial burst of photons, since the thermalized photons are again
able to produce hard photons by Comptonization in the sphere.

In order to be able to study the effect of radially symmetric disturbances
of the accretion disk on the temporal behavior of the emerging
Comptonization spectrum we computed the Green's function for photons
emitted emitted from a ring of radius $r$. For this paper we take $r$ to be
twice the coronal radius. The \emph{initial} seed photon energy
distribution was taken to be a delta-function at a prescribed energy
($40.4$\,eV, $115$\,eV, and $214.4$\,eV for the simulations shown in
Figure~3,
cf. \S\ref{sec:compcon}). Those Comptonized photons, however, that are
subsequently reprocessed in the disk are taken to obey a multicolor
blackbody spectrum with a prescribed maximum temperature (here, $150$\,eV)
and radial temperature dependence $\propto R^{-3/4}$. We chose this
procedure because we wished to consider the effects of Comptonization on
phase lags intrinsic to the seed photons, and furthermore because the
coronal structure in the linear code was \emph{fixed} to that of our best
fit non-linear model as presented in paper I.  For the computation of the
time lag, the pathlength of the photon was integrated from its generation
to the point where the photon was leaving the system. To avoid artificial
phase lags introduced by the size of the system, this latter point was
defined to lie on a sphere encompassing the whole system.

\begin{figure*}
\centerline{
\psfig{figure=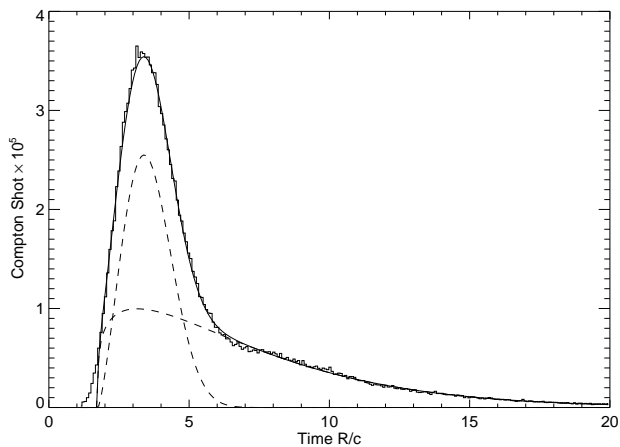,width=0.48\textwidth}
}
\caption
{A Compton shot, for the Comptonization model of paper I, fit with two
  Weibull functions with a common starting time.  The $x$-axis is in
  units of the radial light crossing time of the sphere, and the $y$-axis
  is the number of photons exiting the system
  in a restricted energy band at each time step.}
\label{fig:shot}
\end{figure*}

The propagation of a large number ($7.5 \times 10^5$) of photons was followed
until their statistical weights were very small.  (Nearly all photons had
left the system within $< 30$ light crossing times.)  A typical resulting
output `Compton shot' is shown in Figure~7.
Note that the shot has a narrow component occuring at early times and a
broad component occuring at later times.  This broad component is partly
attributable to reprocessing in the disk; it is the ``echo'' mentioned
above.  Properly normalized, the total shot \emph{is} the transfer
function, $r_{ij}^{[lk]}$, described in eq.~(\ref{eq:tspace}).  Taking
Fourier transforms of such shots, we were able to derive the theoretical
time delays of \S\ref{sec:compcon} and \S\ref{sec:kht}.  We took the
Fourier transforms of analytic fits to the Compton shot profile in order to
avoid spurious power, especially at high Fourier frequencies, due to
numerical counting noise. We found that the combination of two generalized
gamma distributions, specifically two Weibull distributions with a common
start time, were excellent fits to \emph{all} the shot profiles that we
considered.  A sample fit is shown in Figure~7.
We describe these distributions in more detail in the following appendix.

\bigskip
\centerline{B. GENERALIZED GAMMA AND WEIBULL DISTRIBUTIONS}
\bigskip

The probability density function of the generalized Gamma distribution is
given by
\begin{eqnarray}\label{eq:appgamdist}
P_\Gamma(x; \alpha,\beta,\gamma) =
& \frac{\alpha}{\beta} \left [{\gamfn{\frac{1+\gamma}{\alpha}}} \right
  ]^{-1} \left( \frac{x}{\beta} \right)^\gamma \exp \left[ -\left (
  \frac{x}{\beta} \right )^{\alpha} \right ] & {\rm for} ~x>0 \nonumber \\
  & 0 & {\rm for} ~x \le 0
\end{eqnarray}
where
\begin{equation}
\gamfn{x}=\int_0^{\infty} t^{x-1} \exp(-t) ~dt
\end{equation}
is Euler's Gamma function. The generalized Gamma distribution is one of the
most studied probability density functions of statistics since many of the
important non-discrete density functions can be derived from $P_\Gamma$.
For example, $P_\Gamma(x;2,\sqrt{2\sigma},0)$ is the one-sided normal
distribution, and $P_\Gamma(x;1,2,n/2 -1)$ is the $\chi^2$ distribution.
The properties of the generalized Gamma distribution are discussed by
\citey{gran:92a} to whom the reader is referred for a more
extensive discussion\footnote{The notation used here is different from that
  used by \citey{gran:92a}. His notation can be mapped onto the notation
  used here by substituting $a\rightarrow (\gamma+1)/\alpha$,
  $h\rightarrow \alpha$, and $A\rightarrow \beta$.}. Note that the
generalized Gamma distribution used here is a generalization of the Gamma
distribution used by \citey{kazanas:97a}, which has one parameter less. 

In the special case of $\gamma=\alpha-1$ the Gamma distribution is called a
Weibull distribution.  This distribution was first used in 1939 by Waloddi
Weibull as an empirical description for the distribution of the strength of
materials to failure (\cite{weibull:39a}). Since then, the distribution has
had a widespread use in many fields outside of engineering mechanics, e.g.,
to describe the mass distribution of crushed materials (\cite{brown:95a}),
to describe the distribution of the force amplitudes exerted by ocean
waves onto swimming platforms and oil rigs (\cite{gran:92a}),
to describe the distribution of wind speeds to produce building codes
(\cite{whalen:96a}), and many others.

The probability density of the Weibull distribution is  given by
\begin{eqnarray}
P_w(x;\alpha,\beta,x_1) \quad =
    & \left(\frac{\alpha}{\beta}\right)
    \left(\frac{x-x_1}{\beta}\right)^{\alpha-1}
    \exp\left[-\left(\frac{x-x_1}{\beta}\right)^\alpha\right]
    & {\rm for}~ x > x_1 \nonumber \\
    & 0 & {\rm for} ~x \le x_1
\end{eqnarray}
where $\alpha,\beta >0$. In many applications, the ``threshold'' or
``location parameter'' $x_1$ is implicitly set to zero (as has been done
for the generalized Gamma distribution above).  For $\alpha > 1$ the
Weibull distribution looks similar to an asymmetric ``bell curve'', while
for $\alpha<1$ the distribution resembles an exponentially decaying
function. Since $\alpha$ determines the shape of the distribution, it is
often called the ``shape parameter''. The parameter $\beta$ is called the
``scale parameter'' since for a given $\alpha$ the variance of the
distribution is uniquely defined by $\beta$.

The properties of the Weibull distribution are discussed fully in
\citey{gran:92a}.  For our fits of the Compton shots we combined two
Weibull distributions, which were constrained to the same $x_1$.  Thus,
including the absolute normalization of $r_{ij}^{[lk]}$ and the relative
normalization of the two distributions, we had seven fit parameters.  This
yielded excellent fits for all of the Compton shots considered for this
work.


\end{document}